\documentclass{aa}  

\usepackage{graphicx}
\graphicspath{{figs/}}
\usepackage{txfonts}
\usepackage{lipsum}
\usepackage{subcaption}         
\usepackage{lscape}             
\usepackage{placeins}           
\usepackage{natbib}   
\bibpunct{(}{)}{;}{a}{}{,}  
\usepackage{amsmath, amssymb, bm}
\usepackage{unicode-math}
\usepackage{gensymb}

\usepackage{graphicx}
\usepackage{dcolumn}
\usepackage{array}
\usepackage{booktabs}
\usepackage{makecell}
\usepackage{caption}
\usepackage{adjustbox}

\usepackage{etoolbox}
\usepackage{color}
\usepackage{hyperref}
\hypersetup{colorlinks=true, citecolor=blue, urlcolor=blue, linkcolor=blue}
\makeatletter
\let\linenumbers\relax
\let\nolinenumbers\relax
\makeatother

\begin{document}

  \makeatletter
  \renewcommand*\fnmsep{\unskip\hbox{\@textsuperscript{\normalfont*}}}
  \makeatother

\begin{sloppypar}

   \title{Measuring the Collisional Evolution of Debris Clusters in an Asteroid System}


   \author{Yutian Wu\inst{1}\inst{2}
         \and Xiaojing Zhang\inst{2}\thanks{email: zhangxiaojing\_175@126.com}
         \and Chenyang Huang\inst{3}
         \and Yang Yu\inst{1}\thanks{email: yuyang.thu@gmail.com}
         }

   \institute{School of Aeronautic Science and Engineering, Beihang University, Beijing 100191, China             
             \and China Academy of Aerospace System and Innovation, Beijing, China             
             \and School of Aerospace Engineering, Tsinghua University, Beijing 100084, China\\ }

   \date{Received December 8, 2025}

 
  \abstract
   {Rotational instability of rubble-pile asteroids can trigger mass shedding, forming transient debris clouds that may provide the initial conditions for secondary formation in binary systems.}
   {We investigate the dynamical and collisional evolution of a debris cloud numerically generated around a Didymos-like progenitor, as a representative case for the early formation of Dimorphos. The analysis focuses on the growth and structural properties of clusters composed of centimetre- to decimetre-scale particles.} 
   {We perform full-scale simulations of debris evolution around a near-critically rotating asteroid using a cross-spatial-scale approach combined with the discrete element method (DEM). To overcome computational timescale limitations, an equivalent cluster-scale simulation framework is introduced to capture the essential collisional growth processes efficiently. These simulations quantify the efficiency of cluster growth and the structural evoution within the debris cloud.}
   {Our simulations reveal that particles shed from a rotationally unstable asteroid exhibit a consistent migration pattern toward low-geopotential regions, which governs the mass distribution and dynamical structure of the debris cloud. The collisional velocity are well described by a Weibull distribution ($\lambda = 0.0642$, $k = 1.8349$), where low-velocity impacts favor accretion. These collisions enable clusters to grow from centimeter-decimeter scales to meter-sized bodies, developing compact, moderately porous structures ($\Delta I \approx 0.8$, $\phi \approx 0.52$). Collisions between meter-sized clusters do not exhibit a bouncing barrier: low-velocity impacts yield Dinkinesh-like shapes, while moderate velocities promote plastic merging and continued growth.}
   {Our findings suggest that rotational-instability-induced shedding and subsequent collisional accretion provide a viable pathway for secondary formation. Future missions such as Hera, DESTINY+, and Lucy will offer critical opportunities to test these mechanisms through direct observations of debris clouds and dust ejecta.}

   \maketitle

\section{\label{sec:level1-1}Introduction}
Observations indicate that nearly $15\%$ of the near-Earth asteroids (NEAs) are binary systems, and the fraction increases up to $65\%$ for fast-spinning bodies with diameters larger than 300~m \citep{WIMARSSON2024}. These binary systems exhibit notable morphological similarties: the primaries are fast-rotating, top-shaped bodies with equatorial bulges. It is generally accepted that the primary undergoes landslides or mass shedding due to YORP-induced spin-up \citep{yu2018dynamical, pajola2024evidence} or impacts forming a debris cloud around it, from which a secondary accretes through gravitational forces, collisions, and tidal effects \citep{agrusa2024direct, WIMARSSON2024}.

Numerical simulations, laboratory experiments and observations have made effort to investigate debris cloud evolution and secondary formation. Smooth Particle Hydrodynamics (SPH) method has been applied to simulate the landslide-driven debris evolution in the vicinity of a top-shaped asteroid \citep{hyodo2022formation, MADEIRA2023dynamical}, and to investigate the possible origin of (469219) Kamo`oalewa \citep{jiao2022optimal,jiao2024asteroid}. Based on continuum mechanics, SPH can efficiently model large-scale continuous processes and is particularly effective in simulating structural reshaping. However, discrete particle-particle collisions dominate the evolution process in dense debris clouds, introducing mesoscopic complexity that cannot be accurately captured by SPH. To address this limitation, researchers applied Discrete Element Method (DEM) to capture particle-scale interactions in the debris cloud. 

\citet{walsh2008rotational} simulated the moon formation via a step-by-step accumulation of discretely ejected particles due to high-friction angle of asteroid. \citet{agrusa2024direct} used the gravitational N-body code \textit{PKDgrav} to investigate the rapid formation and dynamical evolution of Dimorphos. This process is highly chaotic, and the resulting satellites show a strong preference for a prolate shape. \citet{WIMARSSON2024} combined SPH with gravitational N-body code, \textit{GRAINS}, to simulate the evolution process of angular particles. In their simulations, prolate satellites are formed but are further shaped by tidal interactions with the primary, in contrast to the direct accumulation scenario in \citet{agrusa2024direct}.  

Laboratory experiments have mainly focused on collisions and growth of micron- to centimeter-sized particles \citep{wada2013growth,wada2009collisional}. Collisions between porous aggregates can lead to a wide range of outcomes, from perfect sticking to catastrophic fragmentation. \citet{guttler2010outcome} presented the first comprehensive collision model, and \citet{blum2008growth} provided an in-depth review of the underlying physics, showing that the outcomes depend on impact velocity, aggregate size and size ratio, porosity, and monomer-particle size. 
\citet{birnstiel2024dust} reviewed the possible pathways for dust growth beyond the size barriers toward planetesimals. These include the “lucky particle” scenario, in which small particles colliding at tens of meters per second can accrete onto much larger targets \citep{teiser2009decimetre}, and the “fluffy growth” scenario, in which aggregates with low fractal dimensions and large cross sections experience Stokes rather than Epstein drag, allowing them to grow rapidly enough to overcome the radial drift barrier.
However, such experiments are inherently constrained by the limited duration and scale achievable under microgravity conditions, restricting the particle sizes typically to below a few centimetres \citep{wurm2021understanding,blum2018dust}. 

Observations from the Hubble Space Telescope \citep{jewitt2023dimorphos} and imaging instruments aboard in-situ missions have captured boulders and ejecta fragments located on or near asteroid surfaces. These observations reveal that particles ranging from centimeters to decimeters in size are abundant across asteroid surface \citep{pajola2024evidence}.
The surface of (162713) Ryugu is nearly saturated with small boulders of this size range \citep{michikami2019boulder,michikami2021boulder}, whereas OSIRIS-REx images of (101955) Bennu show that the cumulative surface coverage reaches $50\%$ for particles with sizes of $\mathrm{12-18}$ cm \citep{burke2021particle}. Similarly, Dimorphos is covered by a large fraction of cm-dm particles according to the DRACO scientific camera data during the last five minutes of the DART mission \citep{pajola2024evidence,pojala2023Boulder}.
These size-distributed particles dominate the surface morphology of hundred-meter-scale rubble-pile asteroids.

Despite considerable progress achieved through simulations, experiments, and observations, substantial gaps remain in our understanding of the collisional evolution of debris aggregates.
Numerical studies employing SPH and DEM methods typically concentrate on meter-scale boulders, treating smaller particles (with diameters below 2.5~m) as unresolved “matrix” material.
This approximation limits the mesoscopic characterization of debris cloud evolution and tends to underestimate the structural integrity of forming aggregates \citep{raducan2024physical,raducan2024lessons}.
Laboratory experiments have mainly examined the collisional growth of micron-sized particles. Such fine grains, however, are short-lived in post-shedding environments, as they are rapidly removed by solar radiation pressure \citep{ferrari2022ejecta,li2023ejecta}, and therefore play a negligible role in the long-term evolution of debris clouds.
In contrast, centimeter- to decimeter-sized particles dominate the regolith of rubble-pile asteroids and are expected to constitute the primary components of debris clouds. Yet their collisional behavior and structural evolution remain poorly constrained, owing to the limited spatial resolution and short observational timescales of current observations.
Beyond the asteroid context, recent observations of protoplanetary disks around young stars such as DG Tau and HL Tau have revealed the presence of centimeter-sized pebbles \citep{greaves2025nam}, providing further evidence of their crucial role in the earliest stages of asteroid and planetary system formation.

In this paper, we investigate the full-scale evolution of full-scale debris clouds  generated by rotational instability, using the primary of (65803) Didymos as a representative progenitor. We focus on clusters and conduct a series of DEM simulations to investigate the collisional behavior and structural evolution of aggregates composed of cm-dm particles. 
The results quantify the growth limits of clusters dominated by self-gravity and collisions, and characterize the morphological transition occurring during the cluster growth process. 
Section~\ref{sec:level1-2} briefly introduce methods and equations involoved in this study, including full-scale debris cloud modeling and cluster-scale simulation setup. 
Section~\ref{sec:level1-3} presents the results of simulations, addresses the cluster growth efficiency and structural evolution within the debris cloud and investigates the collisional behavior of survival clusters after cross-scale growth. 
Section~\ref{sec:level1-4} summarizes the main results and conclusions of this study.

\section{\label{sec:level1-2}Method}
\subsection{\label{sec:level2-2.1}Full-scale Debris Cloud Modeling}

The evolution of debris cloud around a near-critically rotating asteroid can be divided into three stages: continous mass shedding, ballistic transport, and the collision-dominated evolution in the debris cloud. We adopted the method developed by \citet{Huang_2024} to simulate the full-scale evolution of derbis cloud shed from the surface of a rotationally unstable progenitor. In this section, we present the principal modeling framework and numerical procedures adopted to simulate the full evolutionary sequence.

\subsubsection{\label{sec:level3-2.1.1}Continuous mass shedding}
Rubble-pile asteroids are commonly modeled with a two-layer structure, with a shallow mantle of loose granular regolith covering a stiffer interior, where spin-up can trigger surface landslides that produce a characteristic top-shaped morphology \citep{sanchez2018rotational,cheng2021reconstructing,walsh2018rubble}. 
Mass shedding on a fast-rotating asteroid can occur through two distinct mechanisms: centrifugal instability and slope failure \citep{yu2018dynamical,Huang2022Longterm}.
We define the unstabel region as the area where the centrifugal forces equals or exceeds the normal components of the gravitational: 
\begin{equation}
\nabla V_\textrm{s} \cdot \hat{\textit{\textbf{n}}} \leq  0 
\label{equnstabel}
\end{equation}
where $\hat{\textit{\textbf{n}}}$ represents the outward unit normal vector of the surface, and $\nabla V_\textrm{s}$ denotes the gradient of the geopotential at the surface. Regions satisfying Eq.~\eqref{equnstabel} are gravitationally unstable, allowing regolith particles to detach or levitate under low-energy perturbations.

Mass shedding can also result from slope failure, which occurs when the local slope angle approaches or exceeds the material's friction angle. The slope angle is defined as the angle between the local geopotential force and the unit normal vector \citep{Huang2021Sandcreep,Huang_2024}:
\begin{equation}
\langle -\nabla V_\textrm{s}, -\hat{\textit{\textbf{n}}}\rangle = {\arctan}{\mu}
\label{eqslpoe}
\end{equation}
where $\mu$ indicates a mean friction coefficient. The steepest descent direction of the geopotential potential $\textit{\textbf{l}}$ is:
\begin{equation}
\textit{\textbf{l}} = -\nabla V_\textrm{s} + (\nabla V_\textrm{s} \cdot {\hat{\textit{\textbf{n}}}})\hat{\textit{\textbf{n}}}
\label{eqsteep}
\end{equation}

To describe the mass shedding generated by rotational instability, we employ a Monte Carlo sampling approach. Debris particles are assumed to follow a nominal power-law distribution of slope $q$: 
\begin{equation}
N_{>d} = N_r d^q
\label{eqsNd}
\end{equation}
where $d$ is the particle size ($d_\textrm{l} < d < d_\textrm{u}$), and $N_r$ is the reference value of number (units: $1 / m^q$). Integrating the mass distribution over the full size range gives:
\begin{equation}
N_r = -\dfrac{6(q+3)}{\pi q \rho} \dfrac{\Delta m}{d_\textrm{u}^{(q+3)} - d_\textrm{l}^{(q+3)}}
\label{eqsNr}
\end{equation}
where $\rho$ is the bulk density of regolith material, $\Delta m$ is the total shedding mass, and $d_\textrm{u}$ and $d_\textrm{l}$ are the upper and lower limits of particle size. 
To ensure statistical representativeness and capture the dynamics across different particle sizes, a discrete set of nodal sizes $d_i$ is assigned across the entire size range, and particles are randomly sampled for each $d_i$. This method identifies regions with high probability of surface instability while quantifies the ejecta mass distribution over the full range of particle sizes.

\subsubsection{\label{sec:level3-2.1.2}Ballistic transport model}
Regolith particles in unstable regions could be shed into orbits through surface collisions and gravitational coupling. We adopt (65803) Didymos as a representative progenitor to model the debris cloud formation, as its equatorial area is dynamically unstable under the current spin period of $2.26$ hr. 
At the early stage of shedding, the debris cloud remains dilute, and mutual collisions can be neglected.
To model the dynamical evolution of the released debris, we track the motion of particles originating from the unstable regions using a ballistic model, with their initial velocities set to zero.

Observations suggest that individual shedding events typically last from several days to weeks \citep{Kleyna2019Gault}, whereas sub-millimeter particles are efficiently removed by solar radiation pressure within about two days \citep{yu2017ejecta,yu2018ejectaII,yu2019expansion}.
Consequently, the motion of the remaining centimeter-sized particles is governed mainly by the asteroid's irregular gravitational field and Coriolis forces in the body-fixed frame. The motion of each particle can thus be expressed as:
\begin{equation}
\ddot{\textit{\textbf{r}}} = -2\omega \times \dot{\textit{\textbf{r}}} -\nabla V(\textit{\textbf{r}})
\label{eqorbit}
\end{equation}
where $\textit{\textbf{r}}$ is the position of the particle in the body-fixed fame, ${(\dot -)}$ and ${(\ddot -)}$ are the first and second time derivatives. $\omega$ is the spin rate of the progenitor, and $V(\textit{\textbf{r}})$ is the geopotential defined as the potential of the centrifugal force and the gravity of the asteroid:
\begin{equation}
V(\textit{\textbf{r}}) = -\frac{1}{2}(\omega \times \textit{\textbf{r}}) \cdot (\omega \times \textit{\textbf{r}}) - \mathrm{G}\sigma \iiint_P{\frac{1}{\textit{\textbf{r}}} dV}
\label{eqV}
\end{equation}
where $P$ represents the polyhedral shape model of the progenitor \citep{werner1996exterior}. $\textrm{G}$ is the gravitational constant, and $\sigma$ is the bulk density of the asteroid.

\begin{figure*}[t]
  \centering
  \includegraphics[width=0.85\textwidth]{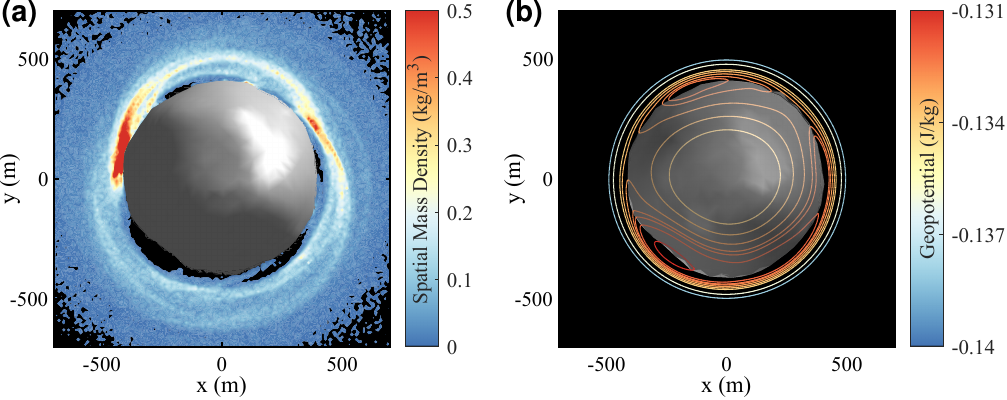}
  \caption{(a) Spatial mass density of the accumulated debris cloud at the end of 2.3-day continuous shedding and ballistic stage, shown in the body-fixed equatorial plane. The black regions depict regions with no detected debris particles. Colored areas mark zones where debris particles are present, and the color scale denotes the local mass density. All regions with mass density exceeding 0.5 are saturated and shown in red.(b) Contour lines of the geopotential around Didymos projected onto the equatorial plane. Different colors indicate distinct geopotential values (unit: $\mathrm{J/kg}$). Values outside the color bar range are shown in the same colors as the respective maximum and minimum values.}
   \label{figgeopotential}
\end{figure*}

In addition to the particles that enter orbit, a certain fraction of the shed debris falls back and re-impacts the asteroid surface. These particles follow a quasi-specular reflection, and the specular reflected velocity is first computed using an inelastic collision model:
\begin{equation}
\left\{ 
  \begin{array}{rcl}
    \textit{\textbf{v}}_\textrm{n}^+  &=&  - c_\textrm{r} \textit{\textbf{v}}_\textrm{n}^-  \\
    \textit{\textbf{v}}_\textrm{t}^+  &=& \textit{\textbf{v}}_\textrm{t}^- -\mu (1+c_\textrm{r}) \Vert \textit{\textbf{v}}_\textrm{n}^- \Vert \dfrac{\textit{\textbf{v}}_\textrm{t}^-}{\Vert \textit{\textbf{v}}_\textrm{t}^- \Vert } \\
  \end{array}  
\right.
\label{eq20}
\end{equation}
where $c_\textrm{r}$ denotes the restitution coefficient. Then, a quasi-specular reflection is introduced using the Cercignani-Lampis-Lord (CLL) method \citep{lord1991some}, where the degree of diffusivity depends on the surface roughness and is governed by the energy accommodation coefficient. In regions that do not satisfy the shedding condition, particles are assumed to accrete onto the surface if their impact velocity falls below below a limit.
Together, continous mass shedding and debris ballistic stage govern the morphology and mass distribution of the debris cloud. The resulting state at the end of a shedding event and is shown in Fig.~\ref{figgeopotential}(a), with contour lines in Fig.~\ref{figgeopotential}(b) denoting the geopotential of Didymos. 
\subsubsection{\label{sec:level3-2.1.3}Collision-Dominated Debris Cloud Evolution}
As the ejected particles accumulate, the debris cloud gradually increases in density and develops heterogeneity. Fig.~\ref{figgeopotential}(a) illustrates the equatorial distribution of particle mass density within the cloud at the end of the ballistic stage, with the densest regions reaching up to $2~\mathrm{kgm^{-3}}$. 
As the particle density rises, interparticle collisions increasingly dominate the evolution, leading to a collapse phase that critically shapes its morphology \citep{yu2018ejectaII,Huang_2024}.
To investigate this process in detail, we employ the \textit{DEMBody} code \citep{cheng2018collision,cheng2021reconstructing,Huang2023Understanding}, which is based on the Soft-Sphere Discrete Element Method (SSDEM) to model interactions between particles under the asteroid's irregular gravitational field.
The total debris mass obtained from the shedding and ballistic stage simulations (Sec.~\ref{sec:level3-2.1.1}-\ref{sec:level3-2.1.2}) is discretized into a non-contact particle cloud that retains the same mass distribution.

The simulated cloud contains 953\,492 centimeter- to decimeter-sized particles. These particles dominate both the mass and the dynamical evolution of the debris cloud several days after the shedding event.
The particle sizes follow a nominal power-law size-frequency distribution (SFD) with a slope of $q=-3.7$, radii ranging from $0.05~\mathrm{m}$ to $0.1~\mathrm{m}$, and a bulk density of $\rho = 3500~\mathrm{kgm^{-3}}$. These parameters are consistent with the SFD inferred from DART mission observations \citep{li2023ejecta,moreno2023characterization}.
To accurately capture the contact mechanics of regolith particles shed from Didymos, the simulations employ realistic physical parameters, which are summarized in Table~\ref{TabelDEM}.

In the DEM simulations, a softened Youngs modulus $Y$ is adopted to allow sufficiently large integration timesteps while maintaining the maximum interparticle overlap below 1\% of the particle radius \citep{zhang2018rotational}. The static and dynamic friction coefficients ($\mu_s$, $\mu_d$), restitution coefficient ($\epsilon$), and shape parameter ($\beta$) are set to typical values for rocky materials, corresponding to an internal friction angle of approximately $40^\circ$ \citep{agrusa2024direct,zhang2022inferring}. Following \citet{raducan2024physical}, the cohesion strength is set to $c = 50~\mathrm{Pa}$, based on comparisons between numerical simulations and observations. The initial spatial and velocity distributions of particles are imported from the ballistic stage result described in Sec.~\ref{sec:level3-2.1.2}.

\begin{table}[h]
    \centering
    \caption{Particle Contact Parameters Used in \textit{DEMBody}.}
    \begin{tabular}{lcc}
        \toprule
        Parameter & Symbol  & Value   \\
        \midrule
        Young modulus    & $Y$  & $5\times10^3~ \mathrm{Pa}$\\
        Poisson's ratio  & $\nu $   & $0.3$  \\
        Static friction coefficient   & $\mu_s$ & $1.0$  \\
        Dynamic friction coefficient & $\mu_d$ & $1.0$\\
        Restitution coefficient & $\epsilon$ & $0.55$\\
        Shape parameter & $\beta$ & $0.8$\\
        Cohesive strength & $c$ & $50~ \mathrm{Pa}$\\
        \bottomrule    
    \end{tabular}
    \label{TabelDEM}
\end{table}

\begin{figure*}[t]
  \centering
  \includegraphics[width=0.85\textwidth]{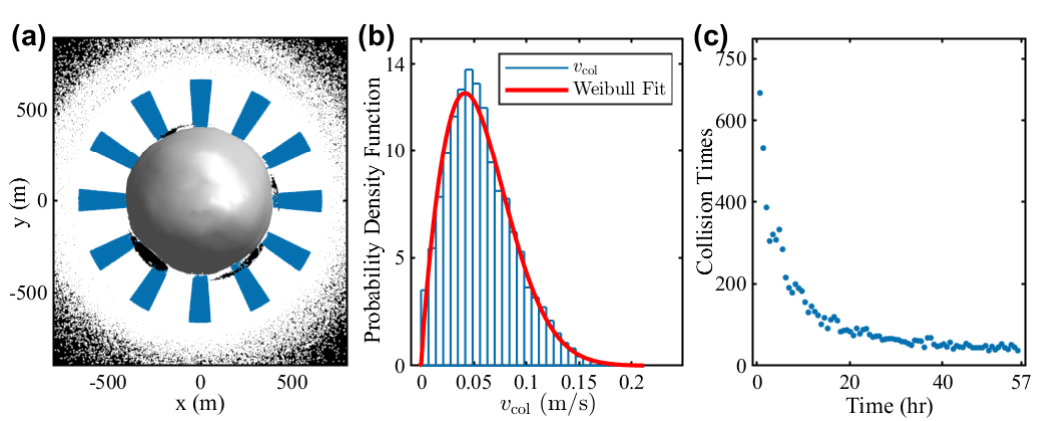}
  \caption{(a) Azimuthal sectors selected within the debris cloud. Blue regions indicate $10$\degree ~wide azimuthal sectors uniformly sampled at orbital radius $r=[360 ~ \mathrm{m},\,660 ~ \mathrm{m}]$. White dots denote simulated particles, while black areas correspond to regions without particles.
  (b) Distribution of the relative collision velocity $v_\mathrm{col}$ extracted from the debris cloud. The red curve shows the fitted Weibull probability density function with a scale parameter $\lambda = 0.0642$ and a shape parameter $k = 1.8349$. (c) Collision counts recorded within the azimuthal sectors.}
  \label{figweibull}
\end{figure*}

To further analyze the collisional behavior within the debris cloud, we extracted representative regions from the full-scale DEM simulation for detailed examination.
Because recording the complete contact history of all 953\,492 particles would exceed the available computational memory, we uniformly selected $12$ sector-shaped regions with radial range $\left[360 ~ \mathrm{m},\, 660 ~ \mathrm{m}\right]$ and central angle $\Delta\theta=10^\circ$ (highlighted in blue in Fig.~\ref{figweibull}(a)) to extract the collision velocities. These sectors correspond to the regions with high mass density in the full-scale debris cloud simulation.
To determine the collision velocities for the subsequent simulations (Sec.~\ref{sec:level2-2.2}), we analyzed the relative collision velocities $v_{\mathrm{col}}$ recorded within the selected sectors. The distribution of $v_{\mathrm{col}}$, spans from $3\times10^{-4}~\mathrm{ms^{-1}}$ to $0.205~\mathrm{ms^{-1}}$, is shown in Fig.~\ref{figweibull}(b). It can be well described by a Weibull probability density function with a scale parameter $\lambda = 0.0642$ and a shape parameter $k = 1.8349$, expressed as:
\begin{equation}
p(v_{\mathrm{col}}) = \dfrac{k}{\lambda}\left(\dfrac{v_{\mathrm{col}}}{\lambda}\right)^{k-1}\exp\left[-\left(\dfrac{v_{\mathrm{col}}}{\lambda}\right)^k\right]
\label{eqweiPDF}
\end{equation}

\subsection{\label{sec:level2-2.2} Simulation Setup of the Cluster-scale Collisional Evolution}
Due to computational constraints, it is impractical to directly simulate the entire formation process from a debris cloud to fully formed satellites at centimeter-scale DEM resolution. To overcome the limited growth efficiency observed in the debris cloud simulations, two complementary cluster-scale DEM experiments were designed: 
(1) Cluster-cluster collisions (Sec.~\ref{sec:level3-2.2.1}), conducted to map the growth efficiency across a wide parameter space of relative collision velocity ($v_\mathrm{col}$) and mass ratio ($\gamma$), thereby identifying the boundary between growth and disruption. 
(2) Sequential particle-cluster collisions (Sec.~\ref{sec:level3-2.2.2}), performed to trace the detailed morphological evolution of a representative cluster undergoing continuous accretion events.

Together, these simulations bridge the gap between full-scale debris evolution and the cluster-scale collisional growth processes, allowing us to evaluate whether clusters can continue to grow under debris cloud conditions. Since these simulations focus on the growth efficiency and morphology of an individual cluster, variations in the gravitational field can be neglected. Only self-gravity and contact between particles are considered.
\subsubsection{\label{sec:level3-2.2.1} Head-on collisions between clusters}
The first set of simulations focuses on head-on collisions between clusters to determine the growth efficiency under varying mass ratios and collision velocities. This experiment aims to identify the transition boundary between growth and disruption, and to evaluate whether clusters formed in the debris cloud can further growth through mutual impacts. In this context, a cluster is defined as a group of particles connected through contact forces. During a cluster-cluster impact, the larger and smaller clusters are referred to as the target and projectile, respectively.

\begin{figure*}[t]
  \centering
  \includegraphics[width=0.8\textwidth]{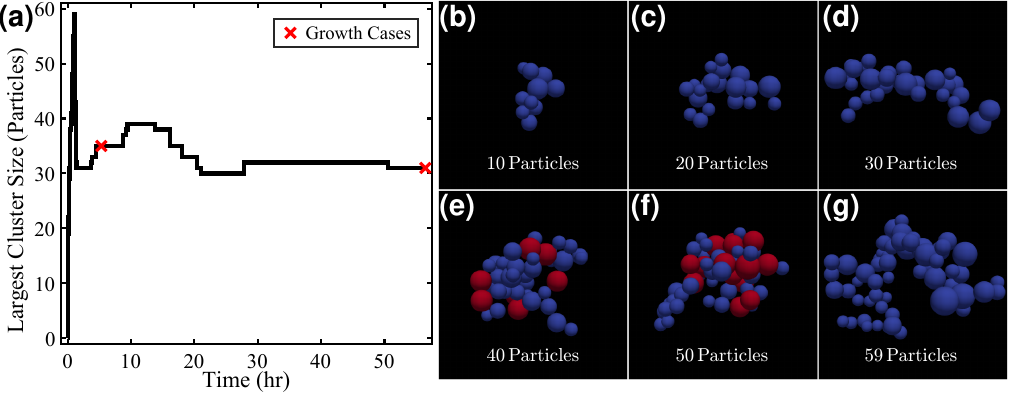}
  \caption{(a) Evolution of the largest cluster size identified at each step in the full-scale simulation, expressed by the number of constituent particles. Red crosses indicate the growth cases used in Sect.~\ref{sec:level3-3.2.2}, whose structures are shown in Fig.~\ref{figthrowclu}(a) and (f). (b-g) Representative clusters sampled during the evolution. Panels (e-f) show supplemented clusters generated using the procedure described in Sec.~\ref{sec:level3-2.2.2}, with red particles indicating the added components.
  }
  \label{figcluMax}
\end{figure*}

Most of the clusters used in this study were directly extracted from the full-scale simulation (Sec.~\ref{sec:level2-2.1}), ensuring physical consistency between the cluster-scale simulations and the full-scale debris environment. The contact parameters are kept identical to those listed in Table~\ref{TabelDEM}.
The clusters extracted from the debris cloud contain between $N=1$ and $60$ constituent particles, with $N=59$ representing the largest cluster formed in the debris cloud (see Fig.~\ref{figcluMax}(a)). However, not every cluster size within this range is formed in the full-scale simulation.
To obtain a more systematic simulation, we supplemented the missing cases by generating additional clusters using the same procedure described in Sec.~\ref{sec:level3-2.2.2}. Representative natural and supplemented clusters are shown in Figs.~\ref{figcluMax}(b-g), with red particles indicating added components.
These clusters exhibit a wide range of morphologies, reflecting the natural diversity of aggregates formed under realistic collisional and gravitational conditions. 

To systematically explore the collisional behavior in the parameter space, all unique cluster pairs are identified, resulting in 1830 distinct combinations. 
Each pair was simulated at 41 collision velocities spanning the full range of $[3\times10^{-4}~\mathrm{ms^{-1}}, \, 0.205~\mathrm{ms^{-1}}]$ measured in the full-scale simulations. To minimize geometric bias, three orthogonal incidence directions are carried out for each case. Simulations are continued until the variation in total system energy between two successive time steps dropped below $10^{-5} ~ \mathrm{J}$, after which the configuration was regarded as a steady post-impact state.
A consistent post-processing routine is then applied to classify each outcome as sticking, mass transfer, or fragmentation.
This approach provides a statistical mapping of the collisional outcomes in the ($v_{\mathrm{col}}$, $\gamma$) parameter space, as discussed in Sec.~\ref{sec:level2-3.2}.

\subsubsection{\label{sec:level3-2.2.2} Procedure for sequential particle-cluster collisions}
The second set of simulations aims to mimic the successive impacts between individual particles and clusters, which represent the most common collision type in a debris cloud.
At the end of full-scale DEM simulation, the low particle number density leads to long collisional timescales and stagnated growth, limiting the exploration of continuous accretion.
To overcome this limitation, we developed a sequential particle-cluster collision algorithm that directly follows the collisional growth of a target cluster through a series of individual impacts.
The algorithm statistically reproduces the velocity and particle size distributions extracted from the full-scale simulation, while substantially reducing computational requirements.

\begin{figure}[htbp] 
   \centering
   \includegraphics[width=0.6\columnwidth]{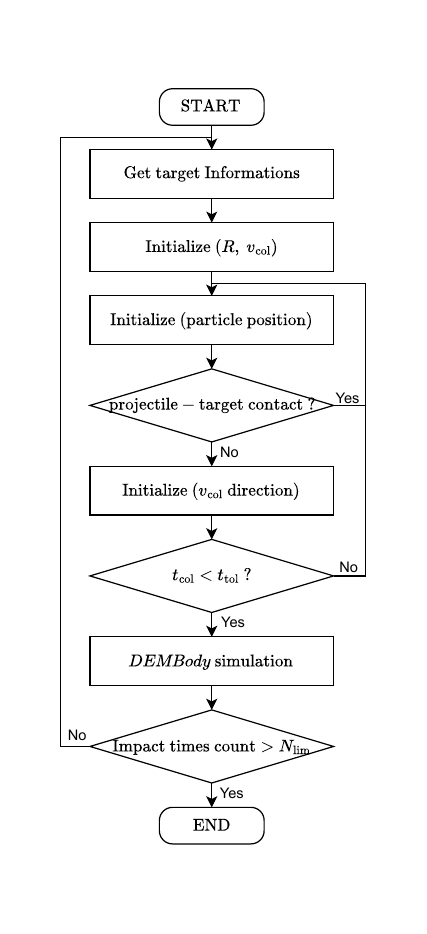}
   \caption{Flow chart of the sequential particle-cluster collision algorithm.}
   \label{figflow}
\end{figure}

Each run involves a single target cluster and one incoming particle, rather than millions of particles as in the full-scale DEM simulation.
This design lowers the computational cost by several orders of magnitude and enables the simulation of hundreds of consecutive impacts to track the structural evolution of clusters.
The overall workflow of the algorithm is illustrated in Fig.~\ref{figflow} and can be summarized as follows:

\noindent \textbf{Step 1: Initialization of particle radius and velocity.}\\
\indent A particle radius $R$ is randomly generated in the range of $\left[0.05 ~ \mathrm{m},\, 0.1 ~ \mathrm{m}\right]$ following a power-law SFD with a slope of $q = -3.7$, consistent with the full-scale simulation. The impact velocity $v_\mathrm{col}$ relative to the cluster mass center follows the Weibull distribution derived in Sec.~\ref{sec:level3-2.1.3} (see Fig.~\ref{figweibull}(b)).

\noindent \textbf{Step 2: Initialization of particle position.}\\
\indent The initial position is randomly chosen outside the cluster to ensure no initial contact.

\noindent \textbf{Step 3: Initialization of velocity direction.}\\
\indent The target cluster is projected onto a plane orthogonal to the vector connecting the particle and the cluster mass center. A random point on this projection is selected as the impact point, and the velocity vector is directed toward it.

\noindent \textbf{Step 4: Collision time estimation.}\\
\indent The relative position and velocity are used to estimate the impact time $t_\mathrm{col}$. If it exceeds a preset threshold $t_\mathrm{tol}$ (i.e., the initial distance is too large), the particle position is resampled (\textbf{Step 2}) with a reduced distance.

\noindent \textbf{Step 5: DEM input generation and iteration.}\\
\indent Once all parameters are accepted, the data are converted into \textit{DEMBody}-compatible input files. After the collision, the final cluster configuration is recorded as the new target for the next iteration. The process continues until the cumulative number of impacts exceeds a predefined limit $N_\mathrm{lim}$. 

Figure~\ref{figcluMax}(e-f) shows examples of cluster growth produced using this procedure.
Post-processing routines were applied to extract structural parameters from each collision for subsequent quantitative analysis.
In summary, the sequential particle-cluster simulation provides a computationally efficient yet physically consistent framework for studying the continuous accretion and structural development of clusters, thereby extending the reach of full-scale debris cloud simulations.

\section{\label{sec:level1-3}Results}
\subsection{\label{sec:level2-3.1}Full-scale Evolution Process in Debris Cloud}
We generate a debris cloud with an initial mass approximately $8\times 10^{-6}$ times that of the progenitor, using the method introduced in Sec.~\ref{sec:level2-2.1}. 
This section presents the results of the full-scale simulation, focusing first on the mass shedding and ballistic stage that lead to the formation and spatial configuration of the debris cloud, followed by its subsequent collisional evolution modeled with the \textit{DEMBody} code.

The debris cloud is produced through a 2.3 days continous shedding and ballistic process. By the end of this phase, the debris cloud exhibits an uneven spatial distribution, showing distinct low-density cavities in the vicinity of Didymos. These cavities correspond to regions of high geopotential $V(\textit{\textbf{r}})$ (Eq.~\eqref{eqV}).
The geopotential contours and mass distribution projected onto the equatorial plane are shown in Fig.~\ref{figgeopotential}, calculated for a rotational period of 2.26~hr. The geopotential field plays a dominant role in shaping the spatial configuration of debris and the motion of individual particles, thus we adopt the Jacobi integral:
\begin{equation}
J = \frac{1}{2}\dot{\textit{\textbf{r}}} \cdot \dot{\textit{\textbf{r}}} + V(\textit{\textbf{r}})
\label{eqJacobi1}
\end{equation}
where the kinetic term $\dot{\textit{\textbf{r}}} \cdot \dot{\textit{\textbf{r}}} \geq 0$.
The contour defined by $C = V(\textit{\textbf{r}})$ represents the zero-velocity curve, which divides the space into two parts: the accessible regions where $C\geq V(\textit{\textbf{r}})$, and the forbidden regions where $C < V(\textit{\textbf{r}})$.
The zero-velocity curve thus constrains the orbital area accessible to each particle \citep{baoyin2010capturing,hirabayashi2013analysis,wu2025dynamic}.

\citet{song2023common} revealed that the equatorial ridge of Didymos mainly corresponds to dynamically unstable regions (see their Fig.~1). Therefore, shed particles repeatedly re-impact the surface and undergo saltation motion \citep{Harris2009shapes}.
Because collisions between the particles and the surface are inelastic, re-impact process causes a monotonic decrease in the Jacobi integral, continuously reducing the accessible regions until the all shedding particles trapped near local minima of the geopotential.
This process is independent of the initial positions of the shed particles. Ultimately, all particles either re-accrete onto the surface or evolve into long-term orbits that do not intersect with the surface of the asteroid.

In the result of shedding stage, the debris cloud naturally develops stable cavities by expelling the particles in the regions of high geopotential, and concentrate around the low-geopotential areas. This spatial heterogeneity creates localized hotspots of particle concentration, which in turn drives further growth through enhanced collisions.
These nonuniform distributions serve as the initial condition for the subsequent DEM simulation. The collisional evolution of the debris cloud was then modeled using the \textit{DEMBody} (see Sec.~\ref{sec:level3-2.1.3}) over a total simulated duration of 56.25 hr.

\begin{figure*}[htbp]
  \centering
  \includegraphics[width=0.85\textwidth]{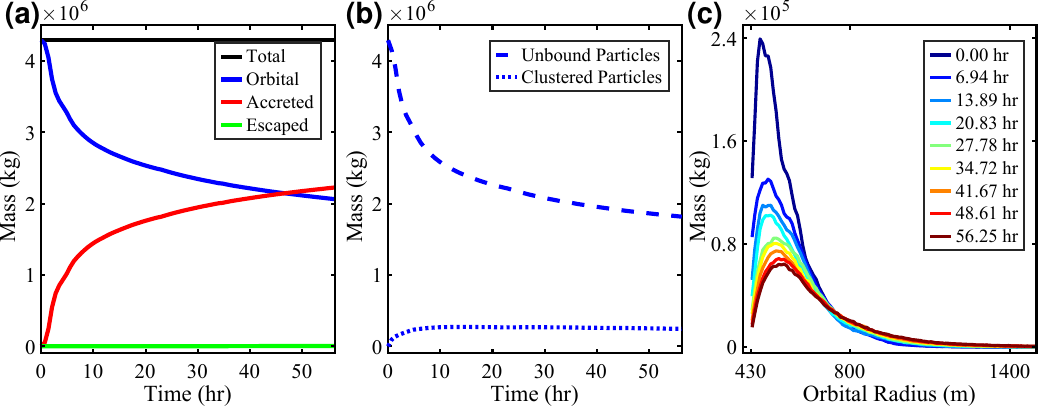}
  \caption{Mass evolution of the debris cloud during the full-scale DEM simulation. 
  (a) Mass evolution of the simulated particles classified into three parts: particles remaining in orbit around the progenitor (blue), accreted onto the progenitor (red), and escaped from the system (green). 
  (b) Mass evolution of unbound particles (dashed line) and clustered particles (dotted line), which together constitute the orbital component.
  (c) Temporal evolution of the radial mass distribution within the debris cloud. Colors correspond to different simulation times as indicated in the legend.}
  \label{figplate}
\end{figure*}

Fig.~\ref{figplate} summarize the spatiotemporal evolution of the debris cloud. The simulated particles can be divided into three parts: (1) accreted onto the progenitor, (2) escaped from the system, and (3) remained in orbit around the progenitor. In Fig.~\ref{figplate}(a), these are shown by the red, green, and blue lines, respectively. The orbital part is further divided into unbound particles and clustered particles, which were individually tracked and quantified in Fig.~\ref{figplate}(b). 

During the first about 10~hr, the debris cloud remained relatively dense, particularly in the low-geopotential regions near Didymos. Collisions became one of the dominant mechanisms governing its evolution. 
Frequent impacts rapidly promoted the formation of clusters, which accounted for approximately $9.5\%$ of the total orbital mass (Fig.~\ref{figplate}(b)). 
Meanwhile, a significant fraction of particles gradually settled back onto the progenitor, resulting in a net reaccreted mass of about $34\%$ of the total debris. The reaccreted material preferentially concentrated in the equatorial region, contributing to the formation of an equatorial ridge \citep{hyodo2022formation}.

In statistics, we find from $t \approx 10$~hr posterior to the shedding event, the debris cloud gradually enters a stable state.
The number of clustered particles becomes stable, and the size of the largest cluster shows little variation (Fig.~\ref{figcluMax}(a); Fig.~\ref{figplate}(b)). 
Both the radial mass distribution (Fig.~\ref{figplate}(c)) and the number of collision times (Fig.~\ref{figweibull}(c)) gradually converge, indicating that the spatial density, radial distribution, and collisional activity change only weakly with time.
Although the system does not become completely static, its overall morphology and statistical properties remain effectively steady, marking a natural transition to the subsequent cluster-scale analysis.

\subsection{\label{sec:level2-3.2}Cluster-scale Growth Conditions in Debris Cloud}
As discussed in Sec.~\ref{sec:level2-3.1}, the full-scale DEM simulations reveal that the debris cloud eventually evolves into a dynamically stable state. Beyond this stage, extending the simulation time yields little additional insight, as collisional growth has essentially ceased and further mass redistribution becomes negligible. It is important to note that our current model captures the evolution following a single shedding event. In reality, however, a rotationally unstable asteroid is likely to experience multiple shedding events, continuously replenishing the debris particles and allowing further growth.

To investigate the intrinsic physics of collisional growth to overcome the limitations of the full-scale model, we therefore perform high-resolution, cluster-scale DEM simulations focused on particle-cluster and cluster-cluster impacts. These localized simulations provide direct access to the cluster-scale growth mechanisms, independent of the global dynamical environment. By systematically varying the impact velocity, mass ratio, and internal structure of the colliding clusters, we quantify how individual collisions drive the collisional evolution of debris clusters under realistic post-shedding conditions.
\subsubsection{\label{sec:level3-3.2.1}Cluster Collisions Mapping in Parameter Space}
We investigate cluster-cluster collisions to evaluate how the mass ratio and collision velocity affect the growth efficiency. The simulation setup follows the methodology described in Sec.~\ref{sec:level3-2.2.1}, using clusters extracted from the full-scale debris cloud simulation. Head-on collisions between all unique cluster pairs are simulated, two growth parameters adapted from \citet{wada2013growth} are used to characterize each collision, i.e. the collisional growth efficiency $f_{\mathrm{gro}}$, and the collisional growth efficiency of the second remnant $f_{\mathrm{2nd}}$. These are defined as
\begin{equation}
f_{\mathrm{gro}}\equiv \dfrac{M_{\mathrm{lar}} - M_{\mathrm{tar}}}{M_{\mathrm{pro}}}
\label{eqfgro}
\end{equation}
\begin{equation}
f_{\mathrm{2nd}} \equiv \dfrac{M_{\mathrm{2nd}} - M_{\mathrm{pro}}}{M_{\mathrm{pro}}}
\label{eqf2nd}
\end{equation}
where $M_{\mathrm{tar}}$ and $M_{\mathrm{pro}}$ denote the masses of the target and the projectile, respectively, with the convention $M_{\mathrm{tar}} \geqslant M_{\mathrm{pro}}$. The quantities $M_{\mathrm{lar}}$ and $M_{\mathrm{2nd}}$ represent the masses of the largest and the second largest remnants after the collision, respectively. 
A collision is said to reach the critical collisional fragmentation velocity $v_{\mathrm{fra}}$ when the target mass remains unchanged ($M_{\mathrm{lar}} = M_{\mathrm{tar}}$). Similarly, the critical collisional transfer velocity $v_{\mathrm{tra}}$ is defined as the collision velocity at which the second remnant retains the original mass of the projectile ($M_{\mathrm{2nd}} = M_{\mathrm{pro}}$).
Based on the definition of growth efficiencies, the collision outcomes are classified into three categories according to the scheme of \citet{hasegawa2021collisional}: \\
1. Growth ($f_{\mathrm{gro}} > 0$): the target gains positive mass through the collision, normalized by $M_{\mathrm{pro}}$.\\
2. Fragmentation ($f_{\mathrm{gro}} < 0$ and $f_{\mathrm{2nd}} <0$): all the remnants are smaller than both the target and the projectile.\\
3. Mass transfer ($f_{\mathrm{gro}} < 0$ and $f_{\mathrm{2nd}} >0$): the second largest remnant exceeds the projectile in mass, while the largest remnant remains smaller than the target.

\begin{figure}[htbp]
   \centering
   \includegraphics[width=0.85\columnwidth]{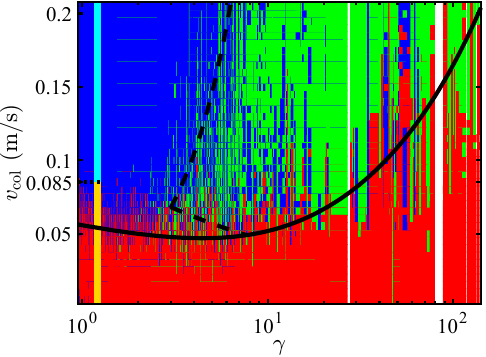}
   \caption{Summary of head-on collision outcomes. The red, blue and green regions represent the collisional growth of the target ($f_{\mathrm{gro}} > 0$), the collisional fragmentation ($f_{\mathrm{gro}} < 0$ and $f_{\mathrm{2nd}} <0$), and the mass transfer from the target to the projectile ($f_{\mathrm{gro}} < 0$ and $f_{\mathrm{2nd}} >0$) respectively. The solid black line denotes the analytical boundary between growth (red) and the other regions (Eq.~\eqref{eqvfra}), while the dashed line marks the transition between fragmentation (blue) and mass transfer (green) region as given by Eq.~\eqref{eqvtra}. The strip at $\gamma = 1.2$ illustrates the outcomes of meter-sized clusters collisions discussed in Sec.~\ref{sec:level3-3.2.3}, where the orange part represents the collisional growth and cyan part represents the collisional fragmentation. Their critical collisional fragmentation velocity $v_{\mathrm{fra,meter}} = 0.085 ~\mathrm{ms^{-1}}$.}
   \label{figcontour}
\end{figure}

Fig.~\ref{figcontour} summarizes the outcomes of collisions with different mass ratios $\gamma = M_{\mathrm{tar}}/M_{\mathrm{pro}}$, and collision velocities $v_{\mathrm{col}}$, as introduced in Sec.~\ref{sec:level3-2.2.1}.
As the growth efficiency is essentially independent of the cluster size \citep{wada2013growth}, we average the results of simulations with identical sets of $\gamma$ and $v_{\mathrm{col}}$ for quantitative analysis.
The red, blue, and green regions in Fig.~\ref{figcontour} correspond to growth, fragmentation, and mass transfer, respectively.

To further determine the boundaries among these categories, we apply the fitting formulation proposed by \citet{hasegawa2021collisional}.
The parameter space can be broadly partitioned into three regimes, divided by the critical velocities for collisional fragmentation ($v_{\mathrm{fra}}$) and mass transfer ($v_{\mathrm{tra}}$), as defined in Eqs.~\eqref{eqvfra} and \eqref{eqvtra}
\begin{equation}
   v_{\mathrm{fra}} \approx \left[(0.055 \times \gamma ^{-0.2})^2 + (0.009 \times \gamma ^{0.63})^2 \right] ^{1/2}
   \label{eqvfra}
\end{equation}
\begin{equation}
   v_{\mathrm{tra}} = \begin{cases}
        0.1 \times \gamma ^{-0.35}, v_{\mathrm{tra}}\lesssim 0.068~\mathrm{ms^{-1}}, \\
        0.013 \times \gamma ^{1.5}, v_{\mathrm{tra}}\gtrsim  0.068~\mathrm{ms^{-1}}.
   \end{cases}
   \label{eqvtra}
\end{equation}

As shown in Fig.~\ref{figcontour} and summarized by Eqs.~\eqref{eqvfra}-\eqref{eqvtra}, the derived $v_{\mathrm{fra}}$ values for $1 \leqslant \gamma \lesssim 12.8$ are lower than the value at $\gamma = 1$ ($v_{\mathrm{fra}}\approx 0.063 ~\mathrm{ms^{-1}}$). The minimum critical collisional fragmentation velocity is $v_{\mathrm{fra,min}}\approx 0.047 ~ \mathrm{ms^{-1}}$. As $\gamma$ increases, $v_{\mathrm{fra}}$ first decreases and then rises sharply. This behavior can be interpreted in terms of the energy partitioning during collisions.

During target-projectile collisions, the kinetic energy is primarily dissipated through plastic deformation, i.e. failure events such as rolling, sliding, twisting and contact formation and breakage within each cluster. Dissipation via elastic deformation is negligible and is therefore omitted in the following analysis \citep{oshiro2025investigating}. 
If the impact energy exceeds the capacity of plastic dissipation, the excess is converted into the kinetic energy of ejected fragments, resulting in fragmentation.

At lower collision energies, collisions result in perfect sticking. 
According to \citet{oshiro2025investigating}, sticking proceeds through compression, transition, and stretching stages. During these stages, most kinetic energy is dissipated through inter-particle failure events. The dissipated energy $E_\mathrm{diss}$ is defined as the difference between the system’s kinetic energy before and after the collision.
For perfect sticking, the dissipated energy in the center-of-mass frame is
\begin{equation}
E_\mathrm{diss} = \dfrac{1}{2} M_\mathrm{r} v_\mathrm{col}^2
\label{eqEdiss}
\end{equation}
where $M_\mathrm{r}$ represents the reduced mass
\begin{equation}
\frac{1}{M_\mathrm{r}} = \frac{1}{M_{\mathrm{tar}}}+\frac{1}{M_{\mathrm{pro}}}
\label{eqMr}
\end{equation}

To investigate the energy partition in detail, we analyze collisions at $v_\mathrm{col} = 0.03~\mathrm{ms^{-1}}$ a velocity that produces exclusively growth outcomes. For each simulation we measure the initial kinetic energy $E_\mathrm{ini}$, the dissipated energy $E_\mathrm{diss}$, the kinetic energy changes $\Delta E_\mathrm{tar}$ and $\Delta E_\mathrm{pro}$, and the number of failure events within each cluster, denoted $N_\mathrm{event,tar}$ and $N_\mathrm{event,pro}$.
In our simulation setups, the initial velocity of projectile is $v_\mathrm{col}$, while the target is set to zero. The initial kinetic energy of the system is
\begin{equation}
E_\mathrm{ini} = \frac{1}{2}M_{\mathrm{pro}}v_\mathrm{col}^2
\label{eqEini}
\end{equation}
Substituting this into Eq.~\eqref{eqEdiss} yields
\begin{equation}
E_\mathrm{diss} = \dfrac{\gamma}{\gamma+1} E_\mathrm{ini}
\label{eqEdissEini}
\end{equation}
 
\begin{figure}[htbp] 
   \centering
   \includegraphics[width=0.85\columnwidth]{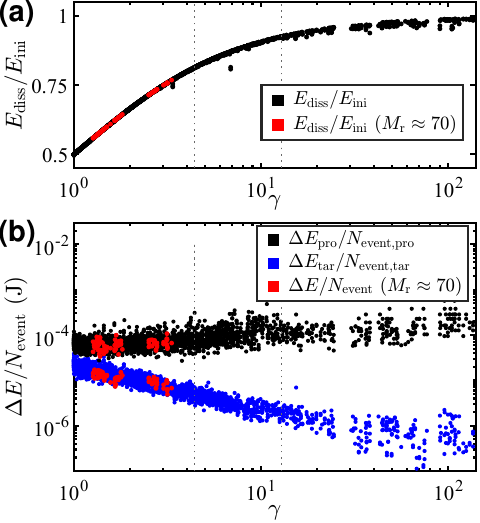}
   \caption{Dissipated kinetic energy and energy partitioning for perfect sticking collisions ($f_\mathrm{2nd}=0$) at $v_\mathrm{col} = 0.03 ~\mathrm{ms^{-1}}$ for different mass ratios $\gamma$. Red symbols indicate a sequence of perfect sticking cases with identical reduced mass $M_\mathrm{r} \approx 70$.
   The vertical dotted lines highlight key mass ratios: $\gamma = 4.4$, where the critical fragmentation velocity reaches its minimum $v_{\mathrm{fra,min}}\approx 0.047~\mathrm{ms^{-1}}$, and $\gamma = 12.8$, where the critical fragmentation velocity $v_{\mathrm{fra}}\approx 0.063~\mathrm{ms^{-1}}$ coincides with that of $\gamma = 1$.
   (a) Dissipated kinetic energy $E_\mathrm{diss}$ of the target-projectile system, normalized by the initial kinetic energy $E_\mathrm{ini}$. 
   (b) Mean dissipated energy per failure event, with $\Delta E_\mathrm{pro}/N_\mathrm{event,pro}$ shown as black points and $\Delta E_\mathrm{tar}/N_\mathrm{event,tar}$ shown as blue points.}
   \label{figEnergy}
\end{figure}

Fig.~\ref{figEnergy}(a) shows the dissipated kinetic energy $E_\mathrm{diss}$ normalized by the initial kinetic energy of the system $E_\mathrm{ini}$.
For collisions with $\gamma = 1$, half of the initial kinetic energy remains after the collision, representing the most efficient energy exchange. 
As $\gamma$ increases, the dissipated energy fraction $E_\mathrm{diss}/E_\mathrm{ini}$ increases rapidly, indicating that a larger fraction of the collisional energy must be dissipated through plastic deformation. 

To isolate the effect of mass ratio alone, we examine a series of perfect sticking cases with fixed reduced mass $M_\mathrm{r} \approx 70$ and identical collision velocity. These collisions share the same total dissipated energy $E_\mathrm{diss}$ and are highlighted as red points in Fig.~\ref{figEnergy}.
It should be pointed out that, for a fixed $M_\mathrm{r}$, increasing $\gamma$ produces a smaller projectile and a larger target simultaneously.
Fig.~\ref{figEnergy}(b) shows that the mean dissipated per failure event exhibits opposite trends in the target and projectile. The target's value $\Delta E_\mathrm{tar}/N_\mathrm{event,tar}$ decreases with $\gamma$ because a larger target can distribute the absorbed energy among more constituent particles. In contrast, $\Delta E_\mathrm{pro}/N_\mathrm{event,pro}$ increases slightly, indicating that the smaller projectile must dissipate comparable energy through fewer contacts. This slow rise suggests that the projectile approaches the upper limit of dissipation achievable through rolling, sliding, and twisting at a single contact.

Because a smaller projectile has reduced capacity for plastic dissipation, it more readily reaches its internal fragmentation threshold as $\gamma$ increases. This explains the initial decline of $v_\mathrm{fra}$ with $\gamma$. At higher mass ratios ($\gamma \gtrsim 4.4$), however, the collision outcome transitions toward cratering-like regime. Although the projectile may be partially disrupted, the massive target can efficiently absorb the kinetic energy of the fragments and often traps them geometrically, making complete sticking more likely. Therefore, the critical fragmentation velocity $v_\mathrm{fra}$ rises again at high $\gamma$.
It should be noted that, due to the number of available cluster sizes is finite, the combinations of target-projectile pairs become increasingly limited at high mass ratios. This limitation leads to numerical gaps in Fig.~\ref{figcontour} around $\gamma \approx 25\,\text{-}\,30$ and $\gamma \approx 80\,\text{-}\,90$.
For $\gamma \gtrsim 66.5$, only particle-cluster collisions occur. In such cases, $f_{\mathrm{2nd}}\geqslant 0$ because individual particles cannot fragment. Consequently, no collisional fragmentation is observed in this regime.

Mass-ratio-dependent behavior consistent with our results has been observed across a wide range of scales and materials. For instance, micron-scale simulations of icy dust conducted by \citet{hasegawa2021collisional} exhibit qualitatively similar collision outcomes (see their Fig.~7). 
At the asteroid scale, the critical disruption energy $Q_D^*$, defined as the specific impact energy required to gravitationally disperse half of the target mass, displays a comparable decrease-increase trend with increasing size \citep{stewart2009velocity}.
Specifically, asteroids with radii smaller than about$100~\mathrm{m}$ resist disruption due to their high intrinsic structural strength. Much larger bodies are difficult to disrupt because of their high gravitational binding energy. In contrast, objects with radii around $100~\mathrm{m}$ are most vulnerable to catastrophic fragmentation. However, despite being easily shattered, fragments from such collisions often reaccumulate into rubble-pile structures held together by mutual gravity.

\subsubsection{\label{sec:level3-3.2.2}Cross-scale Growth Pathways under Sequential Collisions}
The collisional map (Fig.~\ref{figcontour}) show that the velocity range within the debris cloud covers that required for cluster growth. This suggests that collisional growth of clusters can proceed naturally under the typical dynamical conditions of the debris cloud. In realistic environments, multiple shedding events intermittently supply debris particles, thereby sustaining ongoing collisions. Consequently, over sufficient time, survival clusters will inevitably undergo successive collisions and grow to larger scales. Moreover, as a cluster gains mass, its target-projectile mass ratio increases, which further enhances its collisional growth efficiency.
However, computational timescale constraints prevent full-scale debris cloud DEM simulations from being extended long enough to capture sequential cluster growth. To address this, we propose a sequential particle-cluster collision method, detailed in Sec.~\ref{sec:level3-2.2.2}, to investigate the long-term growth behavior of clusters.

\begin{figure*}[t]
  \centering
  \includegraphics[width=0.85\textwidth]{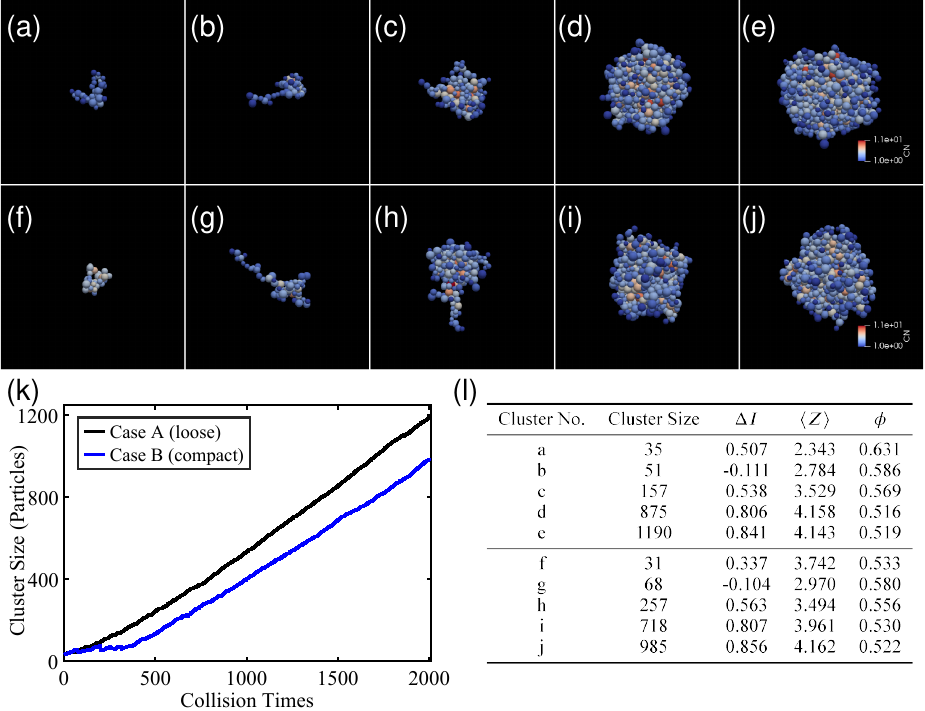}
  \caption{The growth processes of two representative cases in the simulations of particle agglomeration through sequential collisions. The corresponding morphological evolutions are shown in panels (a-j), while their shape parameters are summarized in panel (l).
  Panels (a-e) illustrate the growth sequence of case A (a looser cluster), and panels (f-j) show case B (a more compact cluster). Particle colours indicate their coordination numbers.
  Panel (k) presents the evolution of cluster size (expressed by the number of constituent particles) as a function of collision times, with the black and blue lines corresponding to cases A and B, respectively.
  An animation of the cluster growth processes is available in the online movie.}
  \label{figthrowclu}
\end{figure*}

To study the growth process, we selected two representative clusters. Two of the largest clusters, each containing about 30 particles, were selected as seeds from the debris cloud simulation. These clusters, identified at the early stage and end of the simulation (marked by red crosses in Fig.~\ref{figcluMax}(a)), exhibit markedly different morphologies and packing structures.
Their selection aimed to examine whether initial morphological differences influence subsequent collisional evolution and growth behavior. 
To focus specifically on collisional growth rather than fragmentation, collision velocities were sampled from a Weibull distribution (Fig.~\ref{figweibull}) with an upper limit of $0.08~\mathrm{ms^{-1}}$.
A total of 2000 particle-cluster impacts were simulated for each seed cluster. The growth processes are illustrated in Fig.~\ref{figthrowclu}. Panel (k) shows the evolution of the constituent particle number in each cluster with impact times (the black line corresponds to the cluster shown in panels (a-e), and the blue line corresponds to that shown in (f-j)). 
Both clusters eventually reached approximately $10^3$ constituent particles and adopted nearly spherical shapes with diameters of about $2~\mathrm{m}$. Despite their different initial morphologies, repeated impacts led both clusters to evolve toward similar spherical configurations.

\begin{figure*}[t]
  \centering
  \includegraphics[width=0.85\textwidth]{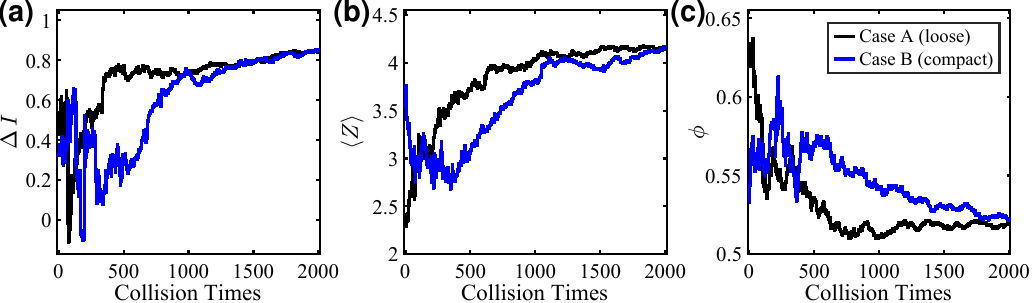}
  \caption{ The evolution of the characterization parameters with collision times. The black lines represent the growth path of case A (a looser cluster, shown in panels Fig.~\ref{figthrowclu}(a-e)), while the blue lines represent case B (a more compact cluster, shown in panels Fig.~\ref{figthrowclu}(f-j)).
  (a) Sphericity index $\Delta I$, where $\Delta I \to 1$ indicates a nearly spherical shapes, and lower values correspond to elongated or irregular clusters. (b) Average coordination number $\langle Z\rangle$, serving as an indicator of the cluster compactness. (c) Macroporosity $\phi$ calculated using a Voronoi tessellation method, also quantifies compactness.}
  \label{figthrowAnalysis}
\end{figure*}

To analyze the structure and morphological evolution of the growing clusters, we introduce three characterization parameters: the sphericity index $\Delta I$, the average coordinate number $\langle Z\rangle $, and the macroporosity $\phi$. Among them, $\langle Z\rangle$ and $\phi$ together describe the internal compactness of the clusters, whereas $\Delta I$ traces the evolution of their overall shape. The changes in these parameters with respect to the collision times are shown in Fig.~\ref{figthrowAnalysis}.

We first examine the sphericity index $\Delta I$, derived from the principal moments of inertia $[I_\mathrm{x},I_\mathrm{y},I_\mathrm{z}]$, quantifies the deviation of a cluster from a perfect sphere: 
\begin{equation}
\Delta I = 1 - \dfrac{\mathrm{max}(I_\mathrm{x},I_\mathrm{y},I_\mathrm{z}) - \mathrm{min}(I_\mathrm{x},I_\mathrm{y},I_\mathrm{z})}{\frac{1}{3}(I_\mathrm{x}+I_\mathrm{y}+I_\mathrm{z}) } 
\label{eqvdI}
\end{equation}
Values of $\Delta I$ close to 1 indicate nearly spherical morphology, whereas lower values correspond to elongated or irregular structures. In both experiments, $\Delta I$ exhibits strong fluctuations during the early growth stage, primarily due to the formation of transient, chain-like branches that are dynamically unstable. 
Furthermore, interparticle friction, cohesion forces, and the non-sphericity of the constituent particles prevent the clusters from achieving a fluid equilibrium shape.
As growth proceeds, $\Delta I$ gradually stabilizes around $0.8$, suggesting that the clusters evolve toward spherical configurations.

To characterizes the compactness of cluster packing, the average coordinate number $\langle Z\rangle$ and the the macroporosity $\phi$ are introduced.
$\langle Z\rangle$ is defined as the mean number of contacts per particle, whereas $\phi$ represents the fraction of void space within a cluster, calculated using a Voronoi tessellation method \citep{okabe2009spatial}. This method divides the space around all particles into individual subdomains, each containing all points closer to that particle than to any other.
During the early growth stage, both $\langle Z\rangle$ and $\phi$ exhibit strong fluctuations due to the transient formation of fragile chain-like branches. As these branches are destroyed through subsequent collisions, the clusters become more compact and spherical. At the same time, $\langle Z\rangle$ stabilizes and $\phi$ provides a reliable measure of macroporosity. These trends indicate that the clusters gradually develop denser internal structures.
Both experiments yield consistent values of $\langle Z\rangle \approx 4.15$ and $\phi \approx 0.52$.

A combined analysis of the structural parameters provides a comprehensive view of cluster morphology and its evolution. 
Clusters with initially loose packing grow more rapidly at early stages due to their higher porosity.
Although this porosity-driven growth rate differs initially, the growth paths converge in later stages. The clusters progress through a sequence of stages: forming chain-like branches (about 60 particles), transitioning to spheres (about 200 particles), and stabilizing at $\Delta I \approx 0.8$ as collisions break unstable branches.
The simulations further demonstrate that, under typical debris cloud collision velocities, survival clusters can continue to grow across size scales and eventually develop into meter-sized clusters. 
These findings suggest that the early collisional environment plays a critical role in determining the initial structural properties of clusters, which subsequently shape the formation process and internal structure of the secondary.

\subsubsection{\label{sec:level3-3.2.3}Collisional Behavior of Meter-scale Clusters}
In Sec.~\ref{sec:level3-3.2.2}, the temporal evolution of clusters within the debris cloud was simulated through sequential particle-cluster collisions. The results indicate that clusters can grow by nearly an order of magnitude under the conditions around Didymos, yielding meter-sized, nearly spherical clusters that retain potential for further growth.
Building on these results, we examine low-velocity, head-on collisions between meter-sized clusters to assess their capacity for continued growth and to understand the processes governing their deformation and possible coalescence.

\begin{figure*}[t]
  \centering
  \includegraphics[width=0.85\textwidth]{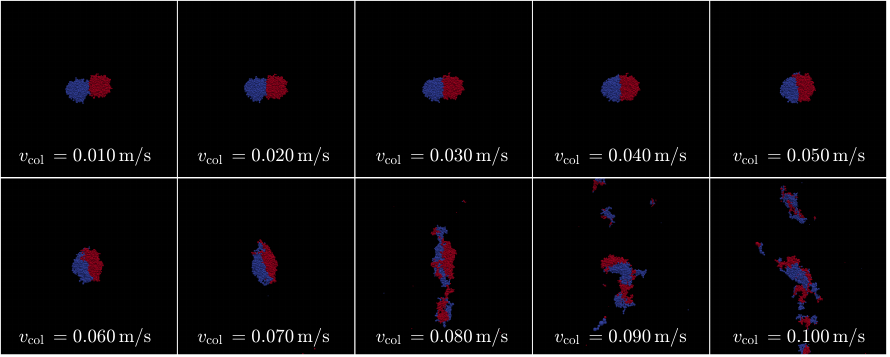}
  \caption{Representative morphologies of collision outcomes between two meter-sized clusters formed in Sec.~\ref{sec:level3-3.2.2}, for relative velocities $v_\mathrm{col}$ ranging from $0.01~\mathrm{ms^{-1}}$ to $0.1~\mathrm{ms^{-1}}$. Particle colors indicate their initial cluster membership. Complete results for collisions within the debris cloud velocity range are available in the Supplementary Material.}
  \label{figbigclu}
\end{figure*}

The two meter-sized clusters obtained from the previous simulations (Fig.~\ref{figthrowclu}(e)(j)) were employed as colliding bodies. Collision velocities were selected according to the velocity distribution within the debris cloud, ensuring realistic dynamical conditions. The outcomes were classified following the criteria defined in Eqs.~\eqref{eqfgro}-\eqref{eqf2nd} and are shown in Fig.~\ref{figcontour} (the strip at $\gamma = 1.2$, where orange indicates collisional growth and cyan denotes fragmentation). Representative results at different collision velocities are presented in Fig.~\ref{figbigclu}. The results reveal four distinct collisional regimes:

\noindent 1. Low-velocity sticking ($v_{\mathrm{col}} \lesssim 0.03~\mathrm{ms^{-1}}$): The two clusters gently adhere upon contact and largely preserve their initial structures, forming a contact binary analogous to asteroid (152830) Dinkinesh I Selam \citep{levison2024contact}.\\
2. Plastic merging ($0.03~\mathrm{ms^{-1}}\lesssim v_{\mathrm{col}} \lesssim 0.055~\mathrm{ms^{-1}}$): The clusters merge into a single, larger cluster with negligible mass loss.\\
3. Damage-enhanced growth ($0.055~\mathrm{ms^{-1}} \lesssim v_{\mathrm{col}} \lesssim 0.085~\mathrm{ms^{-1}}$): The largest post-collision cluster gains mass but becomes morphologically distorted, losing its initial sphericity.\\
4. Fragmentation ($v_{\mathrm{col}} \gtrsim 0.085 ~\mathrm{ms^{-1}}$): Collisions become destructive, and both clusters experience substantial mass loss.

The analysis above, together with Fig.~\ref{figbigclu}, shows that the collisional behavior of meter-sized clusters ($\gamma = 1.2$) qualitatively follows the trend in Fig.~\ref{figcontour}, exhibiting only collisional growth and fragmentation. Quantitatively, the critical collisional fragmentation velocity of meter-sized clusters ($v_{\mathrm{fra,meter}} \approx 0.085~\mathrm{ms^{-1}}$) is significantly higher than that of decimeter-sized clusters ({$v_{\mathrm{fra}} \approx0.054~\mathrm{ms^{-1}}$}) presented in Sec.~\ref{sec:level3-3.2.1}. Notably, the $v_{\mathrm{fra}}$ of decimeter-sized clusters roughly corresponds to the upper velocity limit at which meter-sized clusters still merge into a single, nearly spherical body.

Moreover, the simulations show that the nearly spherical meter-sized clusters undergo only growth or fragmentation with increasing collision velocity, without exhibiting any bouncing behavior.
This contrasts with the behavior of centimeter-scale, fluffy aggregates composed of micron-sized particles, which typically undergo a sticking-bouncing-fragmentation sequence with increasing velocity \citep{schrapler2022collisional}. These results suggest that meter-sized clusters, composed of centimeter- to decimeter-scale particles, possess a highly porous internal structure that efficiently dissipates collision energy within the velocity range of the debris cloud. Such dissipation promotes efficient accretion and allows these clusters to overcome the classical bouncing barrier, facilitating the formation of larger-scale structures.

\section{\label{sec:level1-4}Conclusion}

In this study, we investigated the early stages of secondary formation within debris clouds generated by rotationally unstable asteroids, using full-scale debris cloud model and cluster-scale simulations.
These simulations incorporate realistic asteroid characteristics, accounting for the gravitational field, surface morphology, rotation speed, mass-shedding scale and timescale, particle size distribution, and interparticle contact mechanics.
The full-scale evolution covers a 2.3-day continuous shedding and ballistic stage, followed by a 56.25-hour large-scale, high-resolution DEM simulation. 

Our results show that the progenitor's physical properties yields a unique dynamical environment that governs the mass distribution and cluster evolution within the debris cloud. 
Particles tend to migrate toward low-geopotential regions and enter into orbit. Collision induces growth is the primary process that cluster form in debris cloud. 
The collisional environment can be well characterized by a Weibull distribution ($\lambda = 0.0642$, $k = 1.8349$), whose low-velocity part provides the critical conditions for cluster accretion.

Successive shedding events further sustain a collisional evolution in which both cluster-cluster and particle-cluster impacts occur continuously. 
Cluster-scale simulations reveal the limits of growth within the debris cloud: for mass ratio $\gamma < 4.4$, the critical fragmentation velocity $v_\mathrm{fra}$ decreases with increasing $\gamma$, whereas for $\gamma > 4.4$, it increases. 
This decrease-increase trend indicates a scale-independent pattern, consistent with both laboratory and numerical impact studies spanning from micron-sized particles to hundred-meter bodies.

The sequential particle-cluster collisions enable clusters to grow by nearly an order of magnitude, from centimeter- to decimeter particles to meter-sized clusters. 
Despite differences in their initial configurations, clusters follow a similar evolutionary sequence: chain-like branches emerge when the cluster contains roughly 50 particles, while larger clusters ($\gtrsim 200$ particles) evolve into more compact structure, stabilizing at a sphericity index of $\Delta I \approx 0.8$ and macroporosity $\phi \approx 0.52$.

The subsequent collisions between these meter-sized clusters exhibit distinct regimes. 
Their high-porosity structures provide sufficient internal strength to dissipate collisional energy, avoiding growth stalling like classical bouncing barrier. Instead, their collision outcomes fall into four regimes:
(i) low-velocity sticking ($v_{\mathrm{col}} \lesssim 0.03~\mathrm{ms^{-1}}$), forming Selam-like shapes; (ii) plastic merging ($0.03~\mathrm{ms^{-1}} \lesssim v_{\mathrm{col}} \lesssim 0.055~\mathrm{ms^{-1}}$), producing larger near-spherical clusters; (iii) damage-enhanced growth ($0.055~\mathrm{ms^{-1}} \lesssim v_{\mathrm{col}} \lesssim 0.085~\mathrm{ms^{-1}}$); and (iv) fragmentation ($v_{\mathrm{col}} \gtrsim  0.085~\mathrm{ms^{-1}}$).

Overall, these high-fidelity numerical experiments demonstrate that rotational-instability-induced shedding, followed by collisional accretion, constitutes a prevalent pathway for secondary formation.
The resulting spatiotemporal structure of the debris cloud enables cluster growth across scales, ultimately leading to the emergence of meter-sized structures that may evolve into small binary asteroid systems.
Our findings offer a framework for understanding how rotational shedding and collisional accretion drive secondary formation in asteroid systems. The upcoming missions including Hera, DESTINY+ and Lucy missions, will provide valuable opportunities to verify these mechanisms under diverse dynamical conditions, ranging from impact-generated debris cloud to dust-ejecting asteroid tail and Trojan populations \citep{michel2022esa,simolka2024destiny+,levison2025nasa}. These missions are expected to provide crucial observational evidence on the links between rotational shedding, debris cloud evolution, and secondary formation across diverse small-body populations.

\begin{acknowledgements}
      This work was supported by the National Natural Science Foundation of China (Grant No. 12272018, Y.Y.) and by the Opening Project of Joint Laboratory for Planetary Science and Supercomputing (CSYYGS-QT-2024-17).
\end{acknowledgements}

%
\bibliographystyle{aa} 
\bibliography{refDebris} 

\end{sloppypar}
\end{document}